\documentclass[fleqn,usenatbib]{mnras}

\usepackage[T1]{fontenc}
\usepackage{ae,aecompl}

\usepackage{graphicx}	
\usepackage{amsmath}	
\usepackage{amssymb}	
\usepackage{threeparttable}
\usepackage{multirow}
\usepackage{url}
\usepackage{multicol}
\usepackage{xcolor}

\newcommand{\mum}{\ifmmode{\rm \mu m}\else{$\mu$m}\fi}

\newcommand{\kms}{km~s\ensuremath{^{-1}}}             
  
\newcommand{\chisq}{\ifmmode{\chi^{2} }\else{$\chi^2$}\fi}
\newcommand{\rchisq}{\ifmmode{\chi^{2} }\else{$\chi^2_\nu$}\fi}

\usepackage{orcidlink}
\usepackage{academicons}
\definecolor{orcidlogocol}{HTML}{A6CE39}

\newcommand{\miri}{\texttt{MIRI}}
\newcommand{\jwst}{\textit{JWST}}

\usepackage{newtxtext,newtxmath}

\title[JWST MRS observations of SMP LMC 058]{Observations of the Planetary Nebula SMP LMC 058 with the JWST MIRI Medium Resolution Spectrometer}

\author
[O. C. Jones et al.]{O.~C.\ Jones$^{1}$\thanks{E-mail: olivia.jones@stfc.ac.uk}\orcidlink{0000-0003-4870-5547},
J. \'Alvarez-M\'arquez$^{2}$\orcidlink{0000-0002-7093-1877}, 
G.~C.\ Sloan$^{3,4}$\orcidlink{0000-0003-4520-1044}, 
P.~J. Kavanagh$^{5}$\orcidlink{0000-0001-6872-2358},
I.\ Argyriou$^{6}$\orcidlink{}, 
\newauthor
D.~R.\ Law$^{3}$\orcidlink{0000-0002-9402-186}, 
A.\ Labiano$^{7}$\orcidlink{0000-0002-0690-8824},
P.\ Patapis$^{8}$\orcidlink{0000-0001-8718-3732},
Michael Mueller$^{9}$\orcidlink{0000-0003-3217-5385},
Kirsten L.\ Larson$^{3}$\orcidlink{0000-0003-3917-6460},
\newauthor
Stacey N. Bright$^{3}$\orcidlink{0000-0001-7951-7966},
P.~D.\ Klaassen$^{1}$\orcidlink{0000-0001-9443-0463},
O.~D.\ Fox$^{3}$\orcidlink{0000-0003-2238-1572}$^{3}$,
Danny Gasman$^{6}$\orcidlink{0000-0002-1257-7742}
V.~C.\ Geers$^{1}$\orcidlink{},
\newauthor
Adrian M.~Glauser$^{7}$\orcidlink{0000-0001-9250-1547},
Pierre Guillard$^{10,11}$\orcidlink{0000-0002-2421-1350},
Omnarayani Nayak$^{3}$\orcidlink{0000-0001-6576-6339},
A.\ Noriega-Crespo$^{3}$\orcidlink{0000-0002-6296-8960},
\newauthor
Michael E.\ Ressler$^{12}$\orcidlink{0000-0001-5644-8830},
B.\ Sargent$^{3,13}$\orcidlink{0000-0001-9855-8261},
T.\ Temim$^{14}$\orcidlink{0000-0001-7380-3144},
B.\ Vandenbussche$^{6}$\orcidlink{0000-0002-1368-3109},
\newauthor
Macarena Garc\'{\i}a Mar\'{\i}n$^{3}$\orcidlink{0000-0003-4801-0489}
\\
$^{1}$ UK Astronomy Technology Centre, Royal Observatory, Blackford Hill, Edinburgh, EH9 3HJ, UK \\
$^{2}$ Centro de Astrobiolog\'ia (CSIC-INTA), Carretera de Ajalvir, 28850 Torrej\'on de Ardoz, Madrid, Spain \\
$^{3}$Space Telescope Science Institute, 3700 San Martin Drive, Baltimore, MD 21218, USA \\
$^{4}$Department of Physics and Astronomy, University of North Carolina, Chapel Hill, NC 27599-3255, USA \\
$^{5}$Dublin Institute for Advanced Studies, School of Cosmic Physics, Astronomy \& Astrophysics Section, 31 Fitzwilliam Place, Dublin 2, Ireland \\
$^{6}$Institute of Astronomy, KU Leuven, Celestijnenlaan 200D, 3001 Leuven, Belgium \\
$^{7}$Telespazio UK for the European Space Agency (ESA), ESAC, Spain \\
$^{8}$ETH Zurich, Institute for Particle Physics and Astrophysics, Wolfgang-Paulistr. 27, CH-8093 Zurich, Switzerland \\
$^{9}$Kapteyn Astronomical Institute, University of Groningen, P.O. Box 800, 9700 AV Groningen, The Netherlands \\
$^{10}$Sorbonne Universit\'e, CNRS, UMR 7095, Institut d’Astrophysique de Paris, 98bis bd Arago, \\ 75014 Paris, France \\
$^{11}$Institut Universitaire de France, Minist\'ere de l’Enseignement Supérieur et de la Recherche, 1 rue Descartes, 75231 Paris Cedex 05, France \\
$^{12}$Jet Propulsion Laboratory, California Institute of Technology,4800 Oak Grove Drive, Pasadena, CA 91109 \\
$^{13}$Center for Astrophysical Sciences, The William H. Miller III Department of Physics and Astronomy, Johns Hopkins University, Baltimore, Maryland 21218, USA \\
$^{14}$Princeton University, 4 Ivy Ln, Princeton, NJ 08544, USA
}

\date{Accepted XXX. Received YYY; in original form ZZZ}

\pubyear{2022}

\begin{document}
\label{firstpage}
\pagerange{\pageref{firstpage}--\pageref{lastpage}}
\maketitle


\begin{abstract}
During the commissioning of {\em JWST}, the Medium-Resolution Spectrometer (MRS) on the Mid-Infrared Instrument (MIRI) observed the planetary nebula SMP LMC 058 in the Large Magellanic Cloud.  The MRS was designed to provide medium resolution (R = $\lambda$/$\Delta\lambda$) 3D spectroscopy in the whole MIRI range.   SMP LMC 058 is the only source observed in {\em JWST} commissioning that is both spatially and spectrally unresolved by the MRS and is a good test of {\em JWST's} capabilities.
The new MRS spectra reveal a wealth of emission lines not previously detected in this planetary nebula. 
From these lines, the spectral resolving power ($\lambda$/$\Delta\lambda$) of the MRS is confirmed to be in the range R $=$ 4000 to 1500, depending on the MRS spectral sub-band. In addition, the spectra confirm that the carbon-rich dust emission is from SiC grains and that there is little to no time evolution of the SiC dust and emission line strengths over a 17-year epoch. These commissioning data reveal the great potential of the MIRI MRS for the study of circumstellar and interstellar material.
\end{abstract}





\section{Introduction}\label{intro}

The succession of increasingly powerful mid-infrared spectrographs (e.g., the Short Wavelength Spectrometer (SWS) and the Infrared Spectrograph (IRS) on board the {\em Infrared Space Observatory} and the {\em Spitzer Space Telescope}) launched into space has revolutionised our knowledge of the cool universe \citep[e.g.,][]{Waters1996}.  
For instance the resolving power of the ISO/SWS instrument (R$\sim$1300--2500; \citealt{deGraauw1996}) provided a wealth of detections of line and continuum features never seen before. Whilst the sensitivity of the {\em Spitzer}/IRS \citep{Houck2004} enabled mid-IR spectral observations of a large number of individual objects in external galaxies \citep[e.g.,][]{Zijlstra2006, Kemper2010}.

The Mid-Infrared Instrument (\miri; Wright  et al. ({\em submitted})) on the James Webb Space Telescope (\jwst) includes, in addition to the imager and coronographs, both a low-resolution spectrometer (LRS) covering wavelengths from 5 to 14 $\mu$m \citep{Kendrew15} and a medium-resolution spectrometer (MRS; \citealt{Wells15, Argyriou2023}),  which is an Integral Field Unit (IFU), that has a field of view ranging from $3.2''\times3.7''$ to $6.6''\times7.7''$ (Law  et al. ({\em submitted})) and can spatially resolve spectroscopic data between 4.9 and 27.9 $\mu$m. This is the first time a mid-IR IFU has been deployed outside our atmosphere and through combining and improving upon the best attributes of both the SWS and IRS (namely their spectral resolution and sensitivity) will enable resolved spectroscopic studies of individual stars at the beginning and end of their evolution, diffuse structure in galaxies and planets.

The Large Magellanic Cloud (LMC) is a gas-rich, star-forming, irregular galaxy, which is a satellite of the Milky Way, hosts $\sim$10$^3$  planetary nebulae (PNe) \citep{Reid2010, Reid2014}, and is at a uniform distance of $\sim$50 kpc \citep{Pietrzynski2013}.
In lower metallicity environments like the LMC, which has about half the metallicity of the Milky Way \citep{Westerlund1997, Choudhury2016}, significant dust production is expected to occur in the outflows of asymptotic giant branch (AGB) stars with further processing as these objects become planetary nebula (PNe).  
The gas and dust ejected into the interstellar medium (ISM) by a strong stellar wind from this phase of evolution contains elements synthesised in the stellar interior and dredged up to the surface by convection \citep[e.g.,][]{Karakas2007, Karakas2014}. 
Their chemical composition is expected to primarily depend upon the initial stellar mass and the interstellar elemental abundance at the time the progenitor stars were formed \citep{Kwok2000, Goncalves2014, Kwitter2022}.

As such, the infrared spectra of PNe host a rich variety of features;  forbidden emission lines arising from ionisation from the hot central star \citep[e.g.,][]{Stanghellini2007}, complex organic molecules \citep[e.g.,][]{Ziurys2006}, polycyclic aromatic hydrocarbons (PAHs), and inorganic and organic solids \citep[e.g.,][]{Stanghellini2007, BernardSalas2009, Guzman-Ramirez2011, GarciaHernandez2014} with the frequency of carbonaceous features higher in the LMC than in Galactic PNe. This is likely due to the increased efficiency of third dredge-up (TDU) and the increased C/O ratio at low metallicities \citep{Karakas2002}.
Processing by an external ambient UV radiation field which is stronger in the LMC \citep{Gordon2008} may also affect the circumstellar chemistry. 
For instance, in this metallicity regime where elemental abundances are lower than those in the solar neighbourhood (but maybe comparable to the Galactic anti-centre; \citealt{Pagomenos2018}) dust features like silicon carbide (SiC)  become commonly observed.
Detailed examination of PNe at sub-solar metallicity, therefore, provides a unique insight into chemical abundances and their effect on late-stage stellar evolution, dust production, and the formation of PNe in conditions comparable to those during the epoch of peak star formation in the Universe \citep{Madau1996}. 
Furthermore, due to their compact nature and brightness over a broad wavelength range, PNe are also useful calibration sources \citep[e.g.,][]{Swinyard1996, Feuchtgruber1997, Perley2013, Brown2014}.

SMP LMC 058 was observed by \jwst{} as part of commissioning and calibration activities for MIRI. 
First identified by \cite{Sanduleak1978}, SMP LMC 058 is a carbon-rich planetary nebula (PN) in the LMC, with a heliocentric radial velocity of 278$\pm$7 km s$^{-1}$ \citep{Margon2020}. 
The central star of SMP 058 is a likely C\,{\sc ii}  emitter \citep{Margon2020} with a [WC] spectral type.  
Several dozen very strong, common emission lines of PNe were also detected in its optical spectra \citep{Margon2020}. 
SMP LMC 058 has also been observed with the {\em Spitzer} Infrared Spectrograph (IRS) at both low-resolution (R$\sim$60--127) and high-resolution (R$\sim$600). The Spitzer spectra show SMP LMC 058 has unusual dust chemistry with a strong SiC feature at $\sim$11.3 $\mu$m \citep{BernardSalas2009}  and other associated features, including emission from PAHs at 6--9 $\mu$m, and a shoulder at 18 $\mu$m from an unidentified carrier. However, the Spitzer-IRS data show no clear evidence of fullerenes \citep{Sloan2014}. 
SiC is rarely seen in Galactic PNe, in spite of the higher Si abundance in the Milky Way compared to the Magellanic Clouds \citep{Jones2017b}. Its strength may be due to photoexcitation, or because at a high C/O ratio SiC forms on the surface of carbon grains \citep{Sloan2014}. 

In this paper, we describe the observations and calibration of {\em JWST} MIRI MRS commissioning data of SMP LMC 058  (Section~\ref{obs_cal}). We then present its MRS spectra in Section~\ref{Spec} and determine the resolving power of the MRS in Section~\ref{Resol}.  In Section~\ref{Disc} we identify and analyse the new emission lines and solid-state features detected in this carbon-rich planetary nebula and compare this with {\em Spitzer} IRS data. The potential of the MRS and our conclusions are discussed in Section~\ref{Summ_Conc}.

\section{Observations and Calibrations}\label{obs_cal}

The observations were taken as part of the MIRI MRS commissioning program, program ID 1049 (the commissioning purpose of these observations was point spread function (PSF) characterization).  They use the standard MRS observing template, with 4-point dither patterns optimized for channels 2, 3, and 4 respectively.  Each dither pattern was used twice, in the `positive' and `negative' direction.  Target acquisition was activated, with the science target itself serving as an acquisition target.  
All three bands (SHORT, MEDIUM, LONG)  in all channels were observed in all dithers. Simultaneous MIRI imaging in filter F770W was taken in the dither optimized for channel 2.

A dedicated background observation was taken, employing a 2-point dither optimised for all channels, on a field roughly 3 arcmin away.  The background field was chosen to be relatively clear of astronomical sources based on archival WISE imaging data \citep{Wright2010}.

A total of 45 FASTR1 frames were taken per integration.  In target observations, a single integration was taken per dither point. The background observation had two integrations (to match the total integration time on source, accounting for the use of only a two-point dither on the background).  
The integration time per MRS sub-band and complete dither were therefore 499.5s or roughly 1,500s to cover the entire wavelength range (bands SHORT, MEDIUM, and LONG).
Between the six dithers on-target and the single background, the total integration time was approximately 2.9 hours (6.9 hr including all overheads).

\begin{figure}
\centering
\includegraphics[trim=0.2cm 0cm 0cm 0cm, clip=true,width=0.95\columnwidth]{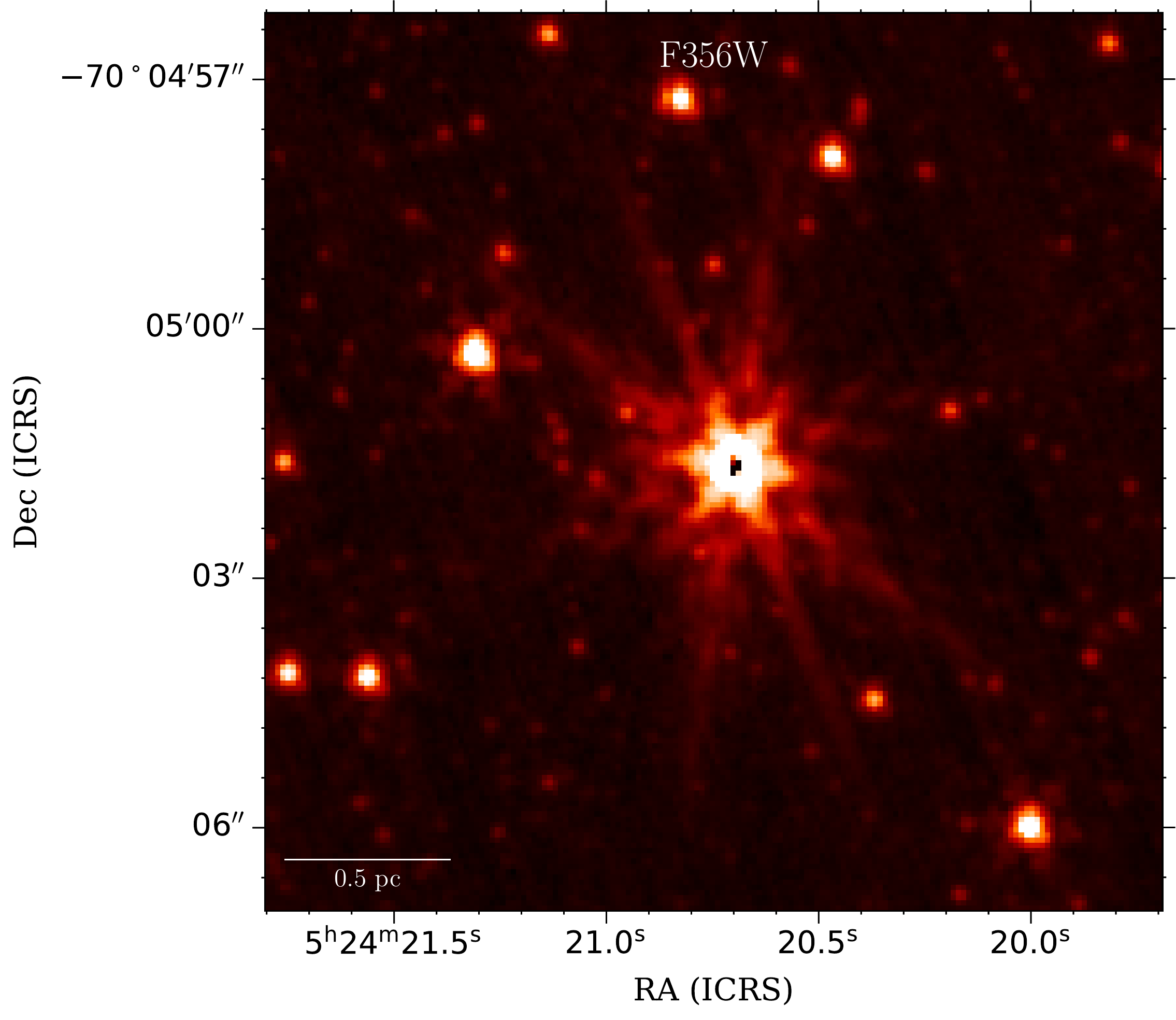}
\caption{NIRCam F356W image of SMP LMC 058 shown in an Asinh stretch. At this spatial resolution (0.063\arcsec) SMP LMC 058 is an unresolved point source.}
\label{Fig:nircamSPMLMC58}
\end{figure}

\begin{figure*}
\centering
\includegraphics[width=\hsize]{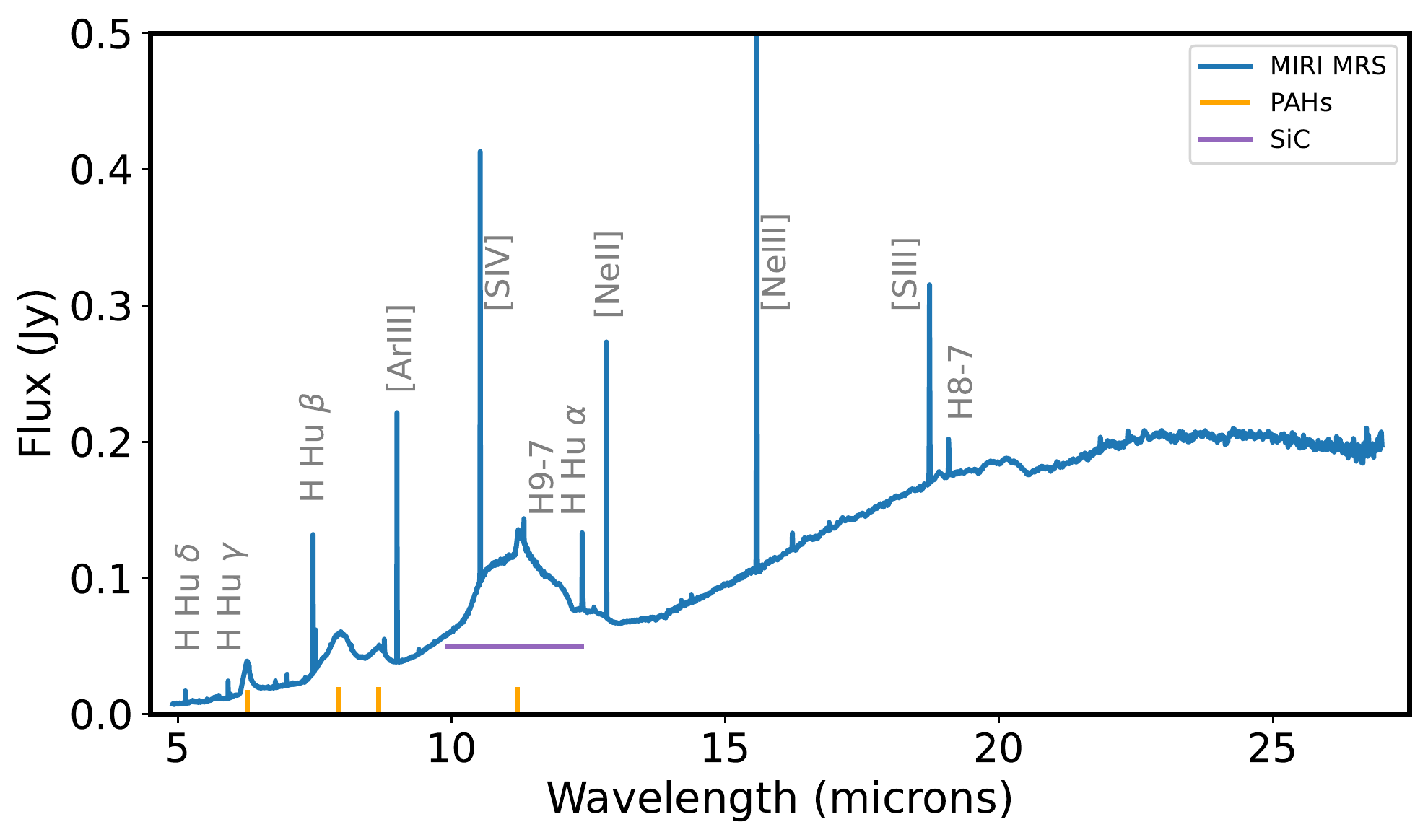}
\caption{The MIRI MRS spectrum of SMP LMC 058.
Numerous emission lines, PAH features and dust features are clearly seen on a rising continuum. These features are much better resolved in the MRS spectra due to the higher spectral resolution.  Artefacts at 12.2 $\mu$m and at $\sim$21 $\mu$m due to a dichroic spectral leak and as a result of stitching channels 4A and 4B are also present.
}
\label{Fig:spec}
\end{figure*}


The MRS observations were processed with version 1.9.5dev of the {\em JWST} calibration pipeline and context 1082 of the Calibration Reference Data System (CRDS).
In general, we follow the standard MRS pipeline procedure (\citealt{MRSpipeline, bushouse_howard_2022_6984366}; and see \citealt{Alvarez-Marquez+22} for an in-flight example of MRS data calibration). The background subtraction has been performed using the dedicated background observation. 
We have generated twelve 3D spectral cubes, one for each of the MRS channels and bands, with a spatial and spectral sampling of 0.13"~$\times$~0.13"~$\times$~0.001~$\mu$m, 0.17"~$\times$~0.17"~$\times$~0.002~$\mu$m, 0.20"~$\times$~0.20"~$\times$~0.003~$\mu$m, and 0.35"~$\times$~0.35"~$\times$~0.006~$\mu$m for channels 1, 2, 3, and 4, respectively. 
The wavelength solution of the data cubes is designed to provide at least 2 samples per spectral resolution element (Law  et al. ({\em submitted})), consequently this Nyquist samples both the spatial PSF and the spectral line spread function (LSF) for each MRS channels.
We have performed 1D spectral extractions individually in each of the MRS cubes using a circular aperture of radius equal to $1.5 \times FWHM (\lambda)$, where $FWHM (\lambda) = 0.3$ arcsec for $\lambda < 8\mu$m and $FWHM (\lambda) = 0.31\times \lambda[\mu m] / 8$ arcsec for $\lambda > 8\mu$m. The selected FWHM\,($\lambda$) values follow the MRS PSF Full Width at Half Maximum (FWHM). NIRCam observation (see Figure \ref{Fig:nircamSPMLMC58}), and MRS observations suggest that SMP LMC 058 is an unresolved source. We use the MRS PSF models (Patapis et al. in prep.) to correct the aperture losses in the 1D spectra. The percentage of flux that is lost out of the selected aperture is 17\% for channel 1 and increases to 30\% in channel 4.

The 12 spectral segments extracted from these cubes were corrected for residual fringing using a post-pipeline spectral-level correction which is a modified version of the detector-level correction available in the JWST calibration pipeline. The residual fringe contrasts are reduced by employing an empirical multi-component sine fitting method \citep[e.g.][]{ref:03KeBeLu}, under the assumption that the pipeline fringe flat correction has reduced fringe contrasts to the point where this multi-component sine approximation is valid (Kavanagh et al., in prep.). 

Finally, each of the 12 individual spectral segments was stitched together to remove minor flux discontinuities. This was done by determining a scaling factor between the median flux (excluding spectral lines) in the overlapping MRS segments; then applying this multiplicative factor to the longer wavelength segments, in turn, to effectively shift the spectrum to match the flux of its neighbouring shorter wavelength segment. This factor was typically on the order of 2--5 per cent, except for channel 4B which required a 9\% scaling. The flux data in the overlapping spectral regions were then averaged. The final stitched spectrum was inspected to ensure there were no remaining discontinuities which may affect the continuum and model fitting.


\section{SMP LMC 058 spectrum}\label{Spec}

\begin{table*}
\caption{Measured central wavelengths, line flux, line widths, and line identification for SMP LMC 058. The systemic velocity was removed prior to calculating the velocity shift of a line. If a line is present in multiple MRS bands, measurements are provided for each individual MRS segment  after a flux scaling factor has been applied.} 
\label{tab:lineIDandstrengths}
\centering
\begin{tabular}{llcccccccc}
\hline
\hline
Band	&	Line 	&	$\lambda_{\rm lab}$	&	$\lambda_{\rm observed}$ &	$\sigma \lambda_{\rm observed}$ &	$\lambda_{\rm offset}$ &	FWHM &	$\sigma$FWHM &	Flux ($\times 10^{-15}$) & $\sigma$ ($\times 10^{-15}$)  \\ 
        &	Identification        & $\mathrm{\mu m}$	&	$\mathrm{\mu m}$ &	$\mathrm{\mu m}$  & km s$^{-1}$ &	km s$^{-1}$	&	km s$^{-1}$	&	$\mathrm{erg\,s^{-1}\,cm^{-2}}$	&	$\mathrm{erg\,s^{-1}\,cm^{-2}}$	\\ 
        
\hline
1S	&	H{\sc i} 23$-$7	&	4.924	&	4.92888	&	0.00036	&	-36.700	&	50.006	&	0.354	&	0.05	&	0.01	\\
1S	&	H{\sc i} 22$-$7	&	4.971	&	4.97548	&	0.00013	&	1.092	&	81.360	&	14.817	&	0.10	&	0.01	\\
1S	&	H{\sc i} 21$-$7	&	5.026	&	5.03059	&	0.00009	&	9.737	&	73.767	&	1.100	&	0.09	&	0.02	\\
1S	&	H{\sc i} 20$-$7	&	5.091	&	5.09597	&	0.00005	&	4.404	&	50.000	&	4.146	&	0.09	&	0.02	\\
1S	&	H{\sc i} 10$-$6	&	5.129	&	5.13323	&	0.00001	&	11.008	&	91.157	&	1.145	&	1.67	&	0.02	\\
1S	&	H{\sc i} 19$-$7	&	5.169	&	5.17383	&	0.00006	&	14.639	&	78.964	&	5.807	&	0.14	&	0.01	\\
1S	&	H{\sc i} 18$-$7	&	5.264	&	5.26835	&	0.00006	&	12.574	&	74.052	&	6.568	&	0.15	&	0.02	\\
1S	&	[Fe {\sc ii}]	&	5.340	&	5.34490	&	0.00009	&	12.537	&	67.901	&	11.809	&	0.06	&	0.01	\\
1S	&	H{\sc i} 17$-$7	&	5.380	&	5.38465	&	0.00005	&	6.445	&	80.560	&	13.568	&	0.19	&	0.02	\\
1S	&	H{\sc i} 16$-$7	&	5.525	&	5.53037	&	0.00003	&	-2.822	&	70.278	&	6.433	&	0.24	&	0.02	\\
1S	&	H{\sc i} 15$-$7	&	5.711	&	5.71656	&	0.00003	&	10.812	&	74.548	&	3.956	&	0.29	&	0.02	\\
1M	&	H{\sc i} 15$-$7	&	5.711	&	5.71681	&	0.00004	&	-2.554	&	83.412	&	5.392	&	0.25	&	0.02	\\
1M	&	H{\sc i} 9$-$6	&	5.908	&	5.91359	&	0.00001	&	5.403	&	92.157	&	1.259	&	2.27	&	0.04	\\
1M	&	H{\sc i} 14$-$7	&	5.957	&	5.96218	&	0.00004	&	9.560	&	88.577	&	3.947	&	0.38	&	0.02	\\
1M	&	[K {\sc iv}]	&	5.982	&	5.98787	&	0.00011	&	-15.674	&	79.335	&	14.445	&	0.09	&	0.01	\\
1M	&	H{\sc i} 13$-$7	&	6.292	&	6.29761	&	0.00004	&	6.860	&	79.420	&	4.690	&	0.47	&	0.03	\\
1L	&	H{\sc i} 12$-$7	&	6.772	&	6.77819	&	0.00001	&	3.630	&	83.820	&	1.616	&	0.59	&	0.02	\\
1L	&	H{\sc i} 21$-$8	&	6.826	&	6.83215	&	0.00013	&	3.098	&	99.481	&	19.478	&	0.09	&	0.02	\\
1L	&	H$_2$(0,0) S(5)	&	6.910	&	6.91573	&	0.00016	&	8.862	&	91.605	&	10.969	&	0.11	&	0.02	\\
1L	&	H{\sc i} 20$-$8	&	6.947	&	6.95288	&	0.00012	&	14.039	&	73.316	&	13.217	&	0.08	&	0.02	\\
1L	&	[Ar {\sc ii}]	&	6.985	&	6.99165	&	0.00001	&	4.671	&	87.664	&	1.052	&	1.19	&	0.01	\\
1L	&	H{\sc i} 19$-$8	&	7.093	&	7.09923	&	0.00009	&	2.620	&	62.794	&	6.562	&	0.09	&	0.01	\\
1L	&	H{\sc i} 18$-$8	&	7.272	&	7.27853	&	0.00013	&	-5.182	&	78.068	&	12.788	&	0.10	&	0.01	\\
1L	&	[Na {\sc iii}]	&	7.318	&	7.32456	&	0.00007	&	-2.741	&	136.659	&	8.951	&	0.49	&	0.03	\\
1L	&	H{\sc i} 6$-$5	&	7.460	&	7.46666	&	0.00001	&	4.907	&	78.844	&	0.318	&	11.59	&	0.05	\\
1L	&	H{\sc i} 8$-$6	&	7.502	&	7.50936	&	0.00001	&	3.950	&	76.638	&	0.636	&	3.90	&	0.07	\\
1L	&	H{\sc i} 11$-$7	&	7.508	&	7.51499	&	0.00002	&	3.142	&	72.973	&	2.467	&	0.72	&	0.02	\\
2S	&	H{\sc i} 15$-$8	&	8.155	&	8.16265	&	0.00011	&	-7.101	&	90.115	&	22.497	&	0.23	&	0.04	\\
2S	&	H{\sc i} 14$-$8	&	8.665	&	8.67260	&	0.00007	&	-2.022	&	63.309	&	15.782	&	0.15	&	0.03	\\
2M	&	H{\sc i} 10$-$7	&	8.760	&	8.76835	&	0.00004	&	-5.273	&	95.058	&	4.011	&	0.92	&	0.04	\\
2M	&	[Ar {\sc iii}]	&	8.991	&	8.99945	&	0.00001	&	9.079	&	95.043	&	0.450	&	20.35	&	0.08	\\
2M	&	H{\sc i} 13$-$8	&	9.392	&	9.40067	&	0.00005	&	2.043	&	81.025	&	5.752	&	0.29	&	0.03	\\
2M	&	H$_2$(0,0) S(3)	&	9.665	&	9.67352	&	0.00023	&	11.069	&	126.523	&	25.282	&	0.19	&	0.03	\\
2M	&	H{\sc i} 18$-$9	&	9.847	&	9.85636	&	0.00020	&	-5.691	&	75.089	&	15.292	&	0.07	&	0.01	\\
2L	&	[S {\sc iv}]  &	10.511	&	10.51997	&	0.00001	&	8.187	&	93.006	&	0.353	&	29.94	&	0.11	\\
2L	&	H{\sc i} 16$-$9	&	10.804	&	10.81342	&	0.00031	&	5.824	&	84.626	&	17.394	&	0.12	&	0.02	\\
2L	&	H{\sc i} 9$-$7	&	11.309	&	11.31891	&	0.00006	&	7.346	&	79.917	&	3.328	&	1.21	&	0.06	\\
3S	&	H{\sc i} 7$-$6	&	12.372	&	12.38329	&	0.00001	&	2.219	&	96.311	&	0.551	&	5.68	&	0.05	\\
3S	&	H{\sc i} 11$-$8	&	12.387	&	12.39839	&	0.00006	&	6.550	&	95.948	&	3.326	&	0.66	&	0.02	\\
3S	&	H{\sc i} 14$-$9	&	12.587	&	12.59901	&	0.00022	&	-6.041	&	87.309	&	12.233	&	0.14	&	0.03	\\
3S	&	[Ne {\sc ii}]	&	12.814	&	12.82546	&	0.00001	&	-0.298	&	93.821	&	0.237	&	17.58	&	0.04	\\
3M	&	H{\sc i} 13$-$9	&	14.183	&	14.19566	&	0.00029	&	12.515	&	73.945	&	14.232	&	0.20	&	0.04	\\
3M	&	[Cl {\sc ii}]	&	14.368	&	14.37954	&	0.00018	&	33.363	&	90.252	&	10.042	&	0.35	&	0.04	\\
3M	&	H{\sc i} 16$-$10	&	14.962	&	14.97505	&	0.00018	&	21.949	&	50.004	&	0.148	&	0.06	&	0.02	\\
3L	&	[Ne {\sc iii}]	&	15.555	&	15.56957	&	0.00002	&	-0.675	&	135.545	&	0.856	&	165.06	&	1.02	\\
3L	&	H{\sc i} 10$-$8	&	16.209	&	16.22345	&	0.00022	&	12.931	&	94.137	&	5.499	&	0.59	&	0.05	\\
3L	&	H{\sc i} 12$-$9	&	16.881	&	16.89658	&	0.00018	&	-5.117	&	88.578	&	7.354	&	0.28	&	0.02	\\
4S	&	[S {\sc iii}]	&	18.713	&	18.72913	&	0.00002	&	19.760	&	137.009	&	0.492	&	11.30	&	0.05	\\
4S	&	H{\sc i} 8$-$7	&	19.062	&	19.07899	&	0.00005	&	9.471	&	148.337	&	3.018	&	2.39	&	0.05	\\
4M	&	[Ar {\sc iii}]	&	21.830	&	21.84986	&	0.00069	&	8.268	&	175.623	&	22.240	&	0.82	&	0.09	\\
4M	&	H{\sc i} 13$-$10+H{\sc i} 11$-$9	&	22.340	&	22.35556	&	0.00054	&	75.494	&	168.816	&	14.574	&	0.65	&	0.14	\\
\hline
\end{tabular}
\end{table*}

Figure~\ref{Fig:spec} shows the extracted spectrum of SMP LMC 058 which exhibits a rich variety of atomic, molecular and solid state features, including PAHs and silicon carbide, characteristic of carbon-rich material, and a strong continuum which rises towards the longest wavelengths. Due to the superior sensitivity and spectral resolution (see Section~\ref{Resol}) of the MRS, the MIRI spectrum of SMP LMC 058 shows features that are not seen in the {\em Spitzer} IRS data (see Section~\ref{Disc}), notably in the number of emission lines detected. 

\begin{figure*}
\centering
\includegraphics[width=\hsize]{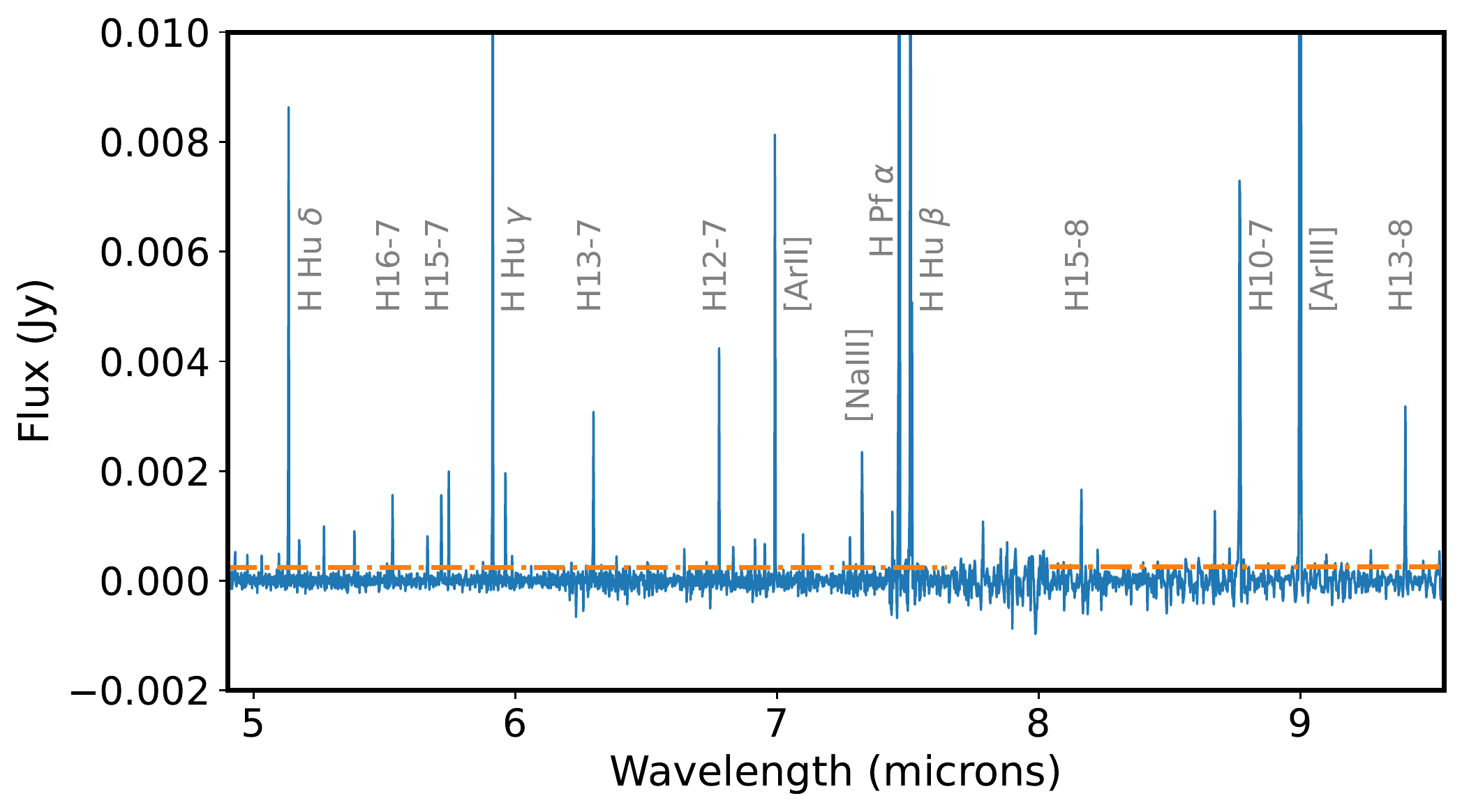}
\includegraphics[width=\hsize]{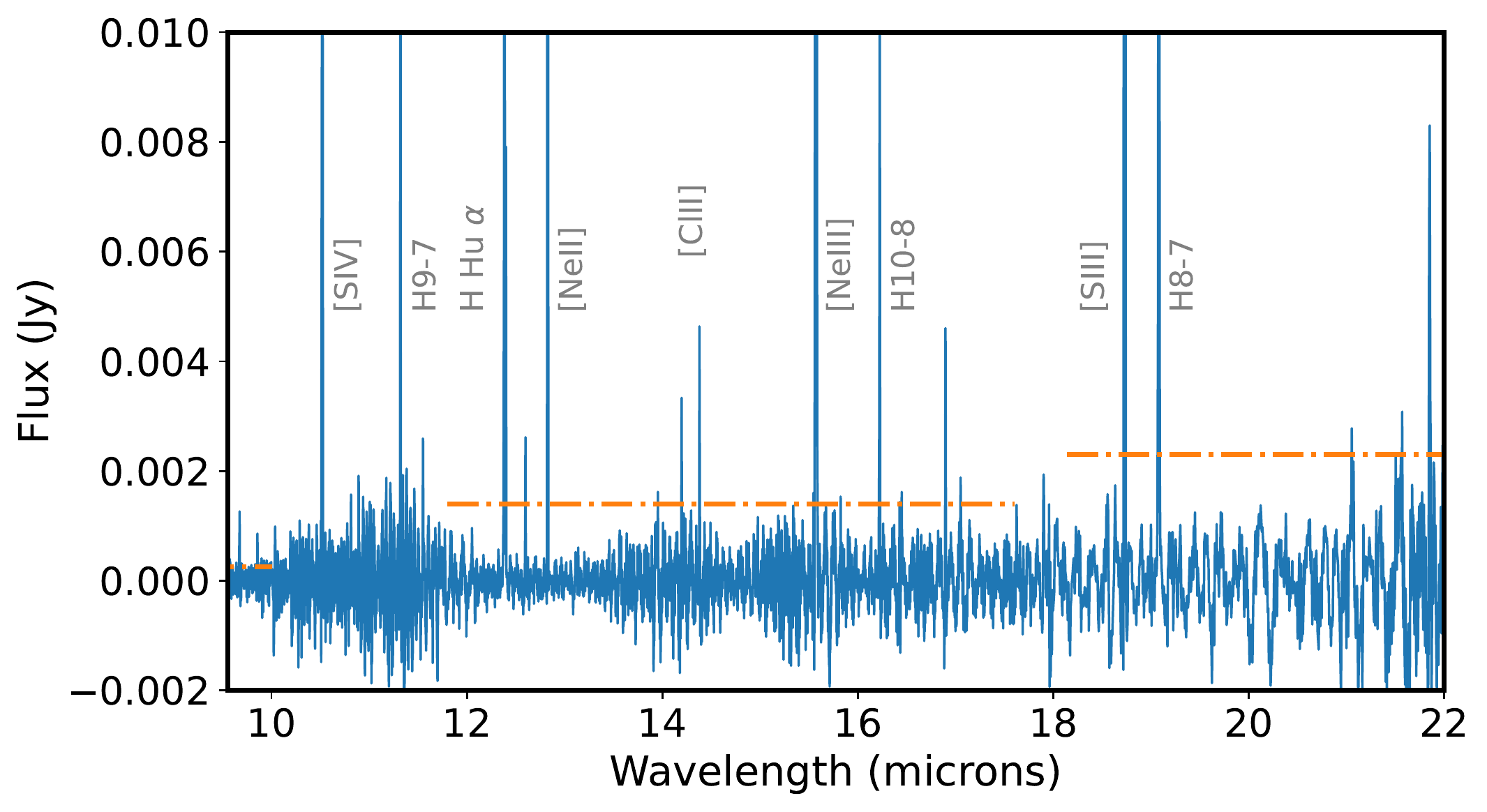}
\caption{Continuum-subtracted MRS spectrum of SMP LMC 058 (where the continuum includes dust and PAH features), highlighting the atomic emission lines. The identification of key species are marked on the spectrum. The top panel shows lines in channels 1 and 2 of the MRS, and the lower panels show channels 3 and 4. 
The orange horizontal dashed lines denote the 3$\sigma$ level above the continuum for each MRS channel, a gap is present where there is a band join between channels or complex continuum emission resulting in an increase in the S/N. The flux axis is truncated to highlight lower contrast lines.}
\label{Fig:spec_contsub}
\end{figure*}

In the spectrum presented here, there is a large amount of fine-structure line emission present, from the strong nebular forbidden lines of [Ar {\sc ii}], [Ar {\sc iii}], [S {\sc iv}], [Ne {\sc ii}], [Ne {\sc iii}], [S {\sc iii}] to weak H recombination lines (H{\sc i}) from the Pfund and Humphreys series, and beyond.
To ensure we measure and identify all the emission lines in the spectra we fit a pseudo-continuum to the broadband spectral features using a piece-wise spline model. Obvious narrow band features were identified and masked in the fitting based on their amplitudes exceeding a threshold value. We used an outlier rejection fitter to flag and ignore any weaker narrow-band features that may compromise the continuum fit. After visual inspection of the fit, it was subtracted to isolate any narrow-band features present.  
Figure~\ref{Fig:spec_contsub} shows the spectrum of SMP LMC 058 after subtraction of the pseudocontinuum from the total spectrum. The spectrum is extremely rich in emission lines. 
In total 51 lines were detected.

\begin{figure}
\centering
\includegraphics[width=\hsize]{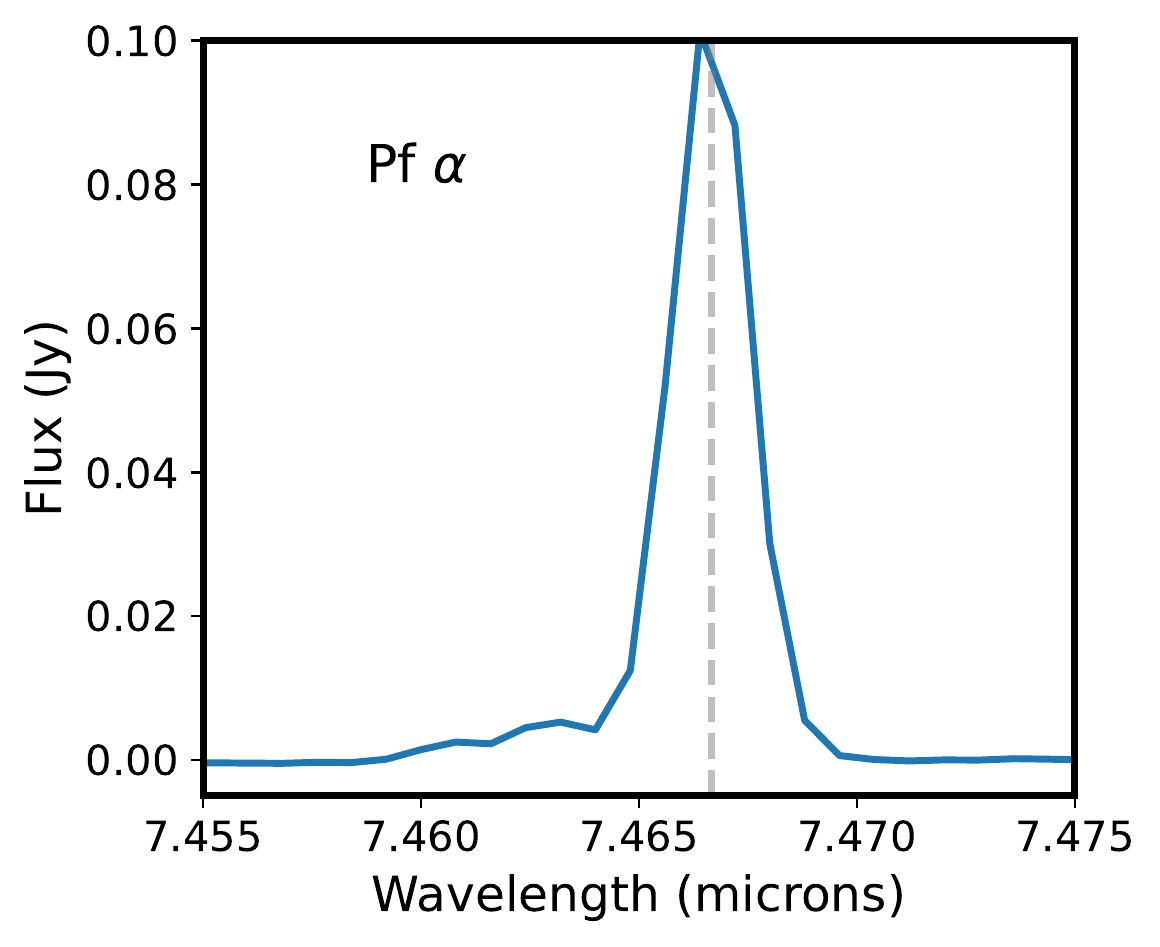}
\includegraphics[width=\hsize]{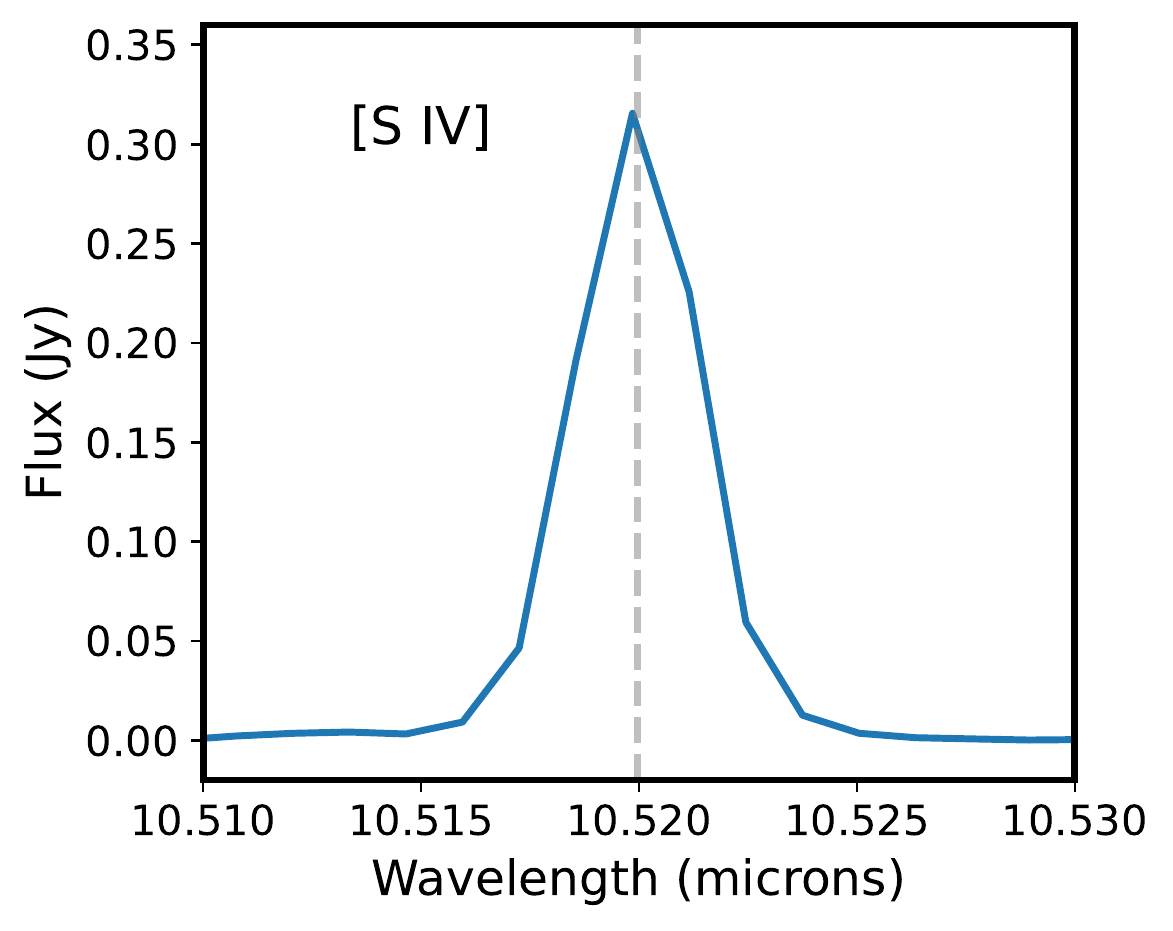}

\caption{Top: The Pf $\alpha$ H {\sc i} emission line profile shows a spatially unresolved main component and a weaker spectrally resolved blue-shifted wing.
Bottom: The [S {\sc iv}] line profile is spectrally unresolved and symmetric. This shape is typical of all the forbidden emission lines in SMP LMC 058. The dashed line marks the lines observed central wavelength.}
\label{Fig:HIlineZoom}
\end{figure}

Using the 12 original MRS segments, we identified and analyzed all detected emission lines with a signal-to-noise ratio (SNR) greater than 3 in the SMP LMC 058 spectra. These individual segments were scaled to remove flux discontinuities prior to measuring the line strengths. Depending on the line profiles (see Figure~\ref{Fig:HIlineZoom}), we performed one-component and two-component Gaussian fits, plus a second-order polynomial to simultaneously fit the continuum and emission line.\footnote{We used the {\sc mpfit} \citep{Markwardt2009} Python routine to perform the fits, the code is publicly available \href{https://github.com/segasai/astrolibpy/tree/master/mpfit}{here}.} The uncertainties on the derived emission line parameters, like the line FWHM, flux, central wavelength, etc, were estimated using a Monte Carlo simulation (following the same methodology as \citealt{Alvarez-Marquez+21,Alvarez-Marquez+22}).  
Systemic velocity shifts were removed using a heliocentric radial velocity of 278 $\pm$ 7 km/s \citep{ReidParker2006, Margon2020}.
Table~\ref{tab:lineIDandstrengths} presents the measured wavelengths and fluxes together with the identification of the mid-IR emission lines in SMP LMC 058 spectra.
The residual scatter in the measured wavelengths is consistent with estimates of the $\sim$ 6\kms\ MRS FLT-5 calibration accuracy (see discussion by \citealt{Argyriou2023}). 
Weak lines are more prevalent at shorter wavelengths in channels 1 and 2 where the  MRS sensitivity is higher and the uncertainties in the flux are better constrained. 

\section{MRS Resolving Power}\label{Resol} 

The resolving power (R) is defined as $\lambda$/$\Delta\lambda$, where $\Delta\lambda$ is the minimum distance to distinguish two features in a spectrum. We define the $\Delta\lambda$ as the FWHM of an unresolved emission line. SMP LMC 058 is the only source observed in the {\em JWST} commissioning datasets that is considered both spatially and spectrally unresolved (with regard to the atomic emission lines) with the MRS, this makes it an excellent target for determining the inflight MRS resolving power. Here, we assume the intrinsic width of the atomic emission lines in SMP LMC 058 to be negligible, as we do not have high-resolution spectroscopy to characterize its intrinsic velocity dispersion. Nearby planetary nebula eject gas with typical velocity dispersions of about 10--51 kms$^{-1}$ \citep{ReidParker2006}. If this is the case for SMP LMC 058, then assuming a velocity of 25 kms$^{-1}$ we might underestimate the MRS resolving power by up to 5\% for channel 1, and up to 1\% for channel 4 \citep[see e.g.,][]{Law2021}. 

\begin{figure*}
\centering
\includegraphics[width=\hsize]{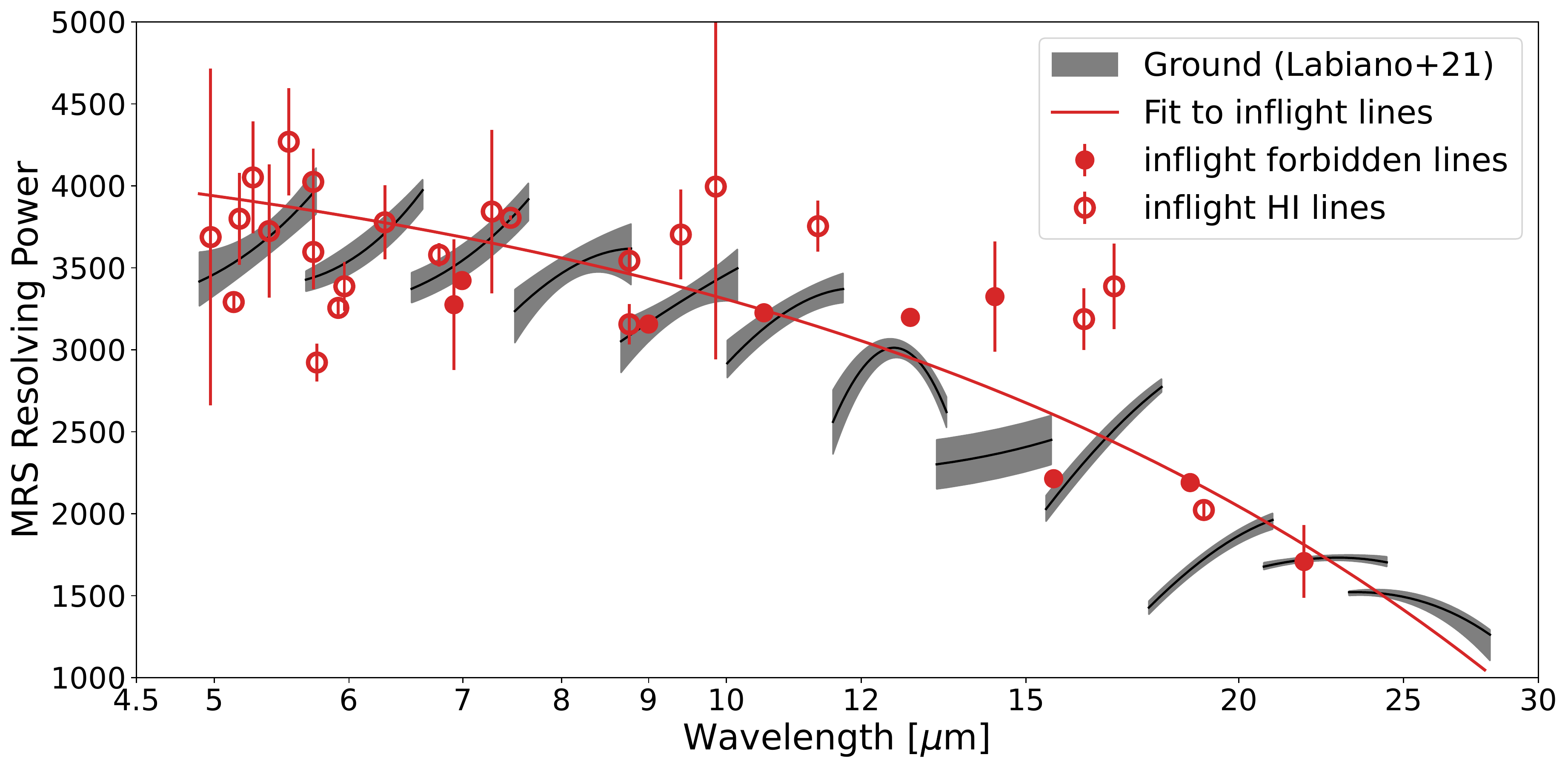}
\caption{Comparison between the ground-based and inflight MRS resolving power. Gray filled area and black line: ground-based MRS resolving power estimates \citep{Labiano21}. Filled red circles: inflight MRS resolving power calculation using forbidden emission lines identified in this paper. Open red circles: inflight MRS resolving power calculation using H{\sc i} emission lines.}
\label{Fig:Resol}
\end{figure*}

The pre-launch MRS resolving power has been established from MIRI ground-based test and calibration campaigns, using a set of etalons which provided lines in all MRS bands. It was determined to be in the range of about 4000 to 1500 \citep{Labiano21}. 
Figure \ref{Fig:Resol} shows the comparison between the ground-based MRS resolving power estimates and the inflight estimates derived from the SMP LMC 058 spectra. The inflight MRS resolving power has been determined using only emission lines with SNR higher than 6, and following the FWHM results obtained in the one- and two-component Gaussian fits (see Section \ref{Spec}). In the case of H{\sc i} emission lines, we used the FWHM of the narrow gaussian component. The errors in the resolving power are, on average, larger for the H{\sc i} emission lines due to the uncertainty in the two-component Gaussian fit. Given the uncertainties, the inflight MRS resolving power agrees with the ground-based estimates, and it presents a trend followed by the equation
\begin{equation}
R(\lambda) = 4603 - 128\times\lambda[\mu m] + 10^{-7.4\times\lambda[\mu m]}.
\end{equation}

The ground-based estimates consider the width of the etalon emission lines to be negligible, which could imply an underestimation of the MRS resolving power by a factor of 10\%. A similar situation is potentially happening with the inflight estimations due to the lack of the intrinsic velocity dispersion of SMP LMC 058. We conclude that the estimations of the ground-based MRS resolving power \citep{Labiano21} are valid, within a 10\% of uncertainty, for the MRS inflight performance. As {\em JWST} observes more sources with spatially and spectrally unresolved spectral lines, their characterisation will provide a more comprehensive understanding of the inflight variations of the resolving power within each of the 12 spectral bands. As of now, the continuous "trend" curve in Figure~\ref{Fig:Resol} presents the state of knowledge of the MRS resolving power.

\section{Discussion}\label{Disc}

\subsection{Emission Lines} 

Short-High and Long-High {\em Spitzer} spectroscopic data of SMP LMC 058 were published in \cite{Bernard-Salas2008}.  A comparison of the MRS line fluxes with those found by \cite{Bernard-Salas2008} is given in Table~\ref{tab:lineFluxcompare}. In general, there is good agreement between our measurements of the forbidden emission line strengths of [S {\sc iv}], [Ne {\sc ii}], [Ne {\sc iii}], [S {\sc iii}]. Furthermore, the high-excitation lines of  [Ar {\sc v}] at 13.10 $\mu$m and [O {\sc iv}] with ionisation potentials of 60 and 55 eV, respectively, are not detected in either spectrum. These lines are excited by high-temperature stars with T$_{\mathrm{eff}\,}$ between 140,000 -- 180,000 K. 
The highest ionisation species in the MRS spectra are [K {\sc iv}] (46 eV) and [Na {\sc iii}] (47 eV), these lines have not been previously detected by {\em Spitzer}. 
Thus, we consider SMP LMC 058 to be a low-excitation source.

Given the superior sensitivity of {\em JWST} \citep{Rigby22} and the MRS, we detect a line at $14.38$ $\mu$m an order of magnitude below the upper-limit of [Ne {\sc v}] reported by \cite{Bernard-Salas2008} which has a rest wavelength of 14.323 $\mu$m. 
The ionisation potential of [Ne {\sc v}] is 97 eV, thus it is unlikely given the absence of other high-excitation lines in the MRS spectra of SMP LMC 058, that this emission is from [Ne {\sc v}], instead, we attribute this line to [Cl {\sc ii}] which has an ionisation potential of 13 eV and a rest wavelength of 14.368 $\mu$m. This identification is consistent with the small wavelength offsets in the current MRS wavelength solution. An upper limit for the [Ne {\sc v}] line at 14.323 $\mu$m is given in Table~\ref{tab:lineFluxcompare}.

\begin{table}
\caption{Comparison of SMP LMC 058 MRS line fluxes with those of \citet{Bernard-Salas2008} taken with the high-resolution modules on the {\em Spitzer} IRS. All line strengths reported by \citet{Bernard-Salas2008} have a 10\% error except for [S {\sc iv}] which has a 10--20\% error. Errors in the MRS flux are $<$1\% and are provided for each line in Table~\ref{tab:lineIDandstrengths}.}
\label{tab:lineFluxcompare}
\centering
\setlength\tabcolsep{4.5pt} 
\begin{tabular}{lcccc}
\hline
\hline
Line & Wavelength  & MRS Flux & {\em Spitzer} Flux  & Ionisation \\   
    &  (Rest)   & $\times10^{-15}$ &   $\times10^{-15}$ & potential \\  
 &  $\mathrm{\mu m}$ &  $\mathrm{erg\,s^{-1}\,cm^{-2}}$ &  $\mathrm{erg\,s^{-1}\,cm^{-2}}$  & (eV) \\ 

\hline
$[$S {\sc iv}]   & 10.511  &  29.99    &  29.2 &   35   \\ 
$[$Ne {\sc ii}]  & 12.814  &  17.58    &  20.6 &    22    \\  
$[$Ar {\sc v}]   & 13.099  &  $<$0.029 &  $<$2.4  &   60  \\  
$[$Ne {\sc v}]   & 14.323  &  $<$0.005 &  $<$3.8 &     97 \\  
$[$Ne {\sc iii}] & 15.555  &  165.06   &  200.6 &      41 \\    
$[$S {\sc iii}]  & 18.713  &  11.30    &  11.0   &       23 \\   
$[$O {\sc iv}]   & 25.883  &  $<$0.23  &  $<$21.6 &   55 \\   
\hline
\end{tabular}
\end{table}

As seen in Table~\ref{tab:lineIDandstrengths} higher ionisation potential species expand at a lower velocity than the lower ionisation potential species \citep[e.g.,][]{ReidParker2006}. This is due to ionisation occurring at a greater distance from the centre of the PN where velocities are greater and can cause lower excitation species to expand to larger radii in the PN. 

Hydrogen recombination lines are abundant in the spectrum of SMP LMC 058, all are new detections.
H{\sc i} emission lines more closely trace the ionized regions, compared to molecular hydrogen. 
As shown in Figure~\ref{Fig:HIlineZoom}, the line profiles of the bright H{\sc i} emission lines are asymmetric, exhibiting a blue tail, whereas the forbidden emission lines present symmetric unresolved profiles. This difference in behaviour is seen even when the lines have a similar SNR, thus unlikely due to a calibration artefact.
H{\sc i} emission lines are composed of a spectrally unresolved main component containing the majority of the line flux ($>$95\%), and a spectrally resolved blue-shifted component possibly  due to thermal broadening \citep{Chu1984} or from condensation outside the main core which may be evident as a marginally resolved  envelope like structure in the MRS cube at 7.466$\mu$m.  

\begin{figure*}
\centering
\includegraphics[trim=0.05cm 0.05cm 0cm 0cm, clip=true,width=\hsize]{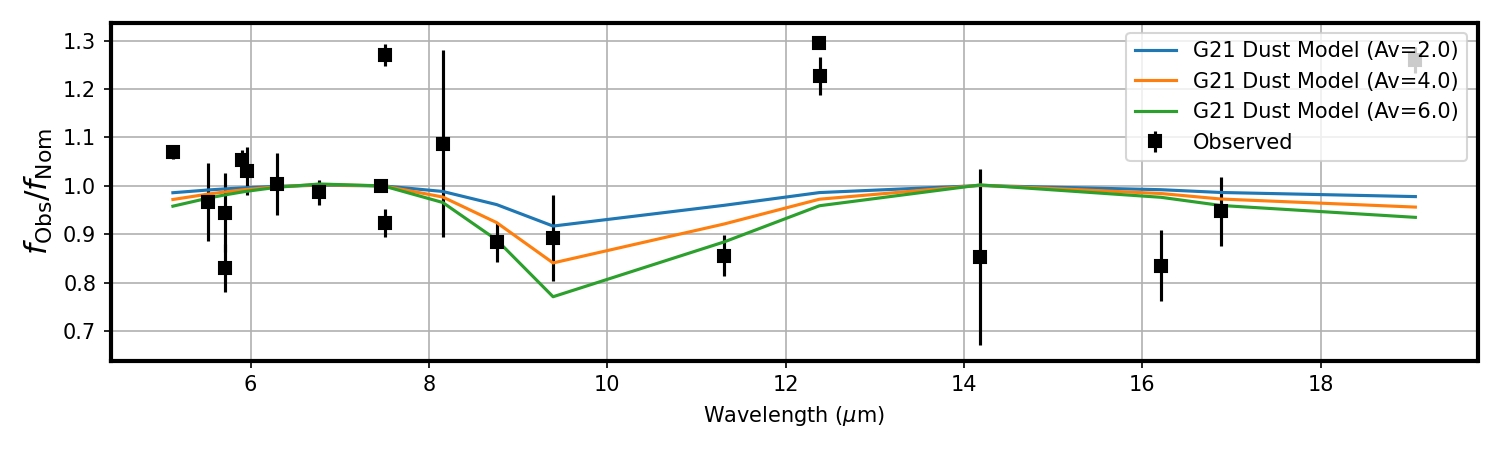}
\caption{Comparison of the nominal H{\sc i} line ratios f$_{\rm Norm}$ from \protect\cite{Hummer1987} and our observed ratios f$_{\rm Obs}$ for SMP LMC 058. Models of dust extinction as a function of wavelength for a range of $A_V$  taken from \protect\cite{Gordon2021} are also shown.
}
\label{Fig:ExtingtionfromHI}
\end{figure*}

Hydrogen recombination lines can be used as a diagnostic for the circumstellar envelope as the line strengths are sensitive to the density of the emitting gas. Low-density gas ($100 < N_e < 10^4$~cm$^{-3}$ and $5000 < T_e < 10^4$~K ) produces optically thin lines characterised by Case B recombination \citep{Baker1938, Storey1995}. 
We compute the observed H{\sc i} flux ratios, normalised to the H{\sc i} 6--5 line, and compare the observed flux ratio to the expected intrinsic flux ratio from \cite{Hummer1987} assuming Case B recombination. In Figure~\ref{Fig:ExtingtionfromHI} we show these values as a function of wavelength compared to the differences that would be expected due to dust extinction following the mid-IR extinction curves of \cite{Gordon2021}.  While there is marginal evidence for a dip in line intensity due to absorption at $\sim$10-micron, the observational uncertainties are too large to allow us to reliably distinguish between extinctions in the range $A_V = 2-6$.

Two molecular hydrogen lines (H$_2$) have been detected in the MRS data,  the ortho-H$_2$ $v =$ 0--0 S(3) and S(5) lines. The S(1) line at 17.055 $\mu$m may also be present, although this is not easily discerned above the continuum and we do not measure its flux. The S(3) and S(5) rotational line emission probably originate from irradiated, and perhaps also shocked, dense molecular clumps, torus structures \citep[e.g.,][]{Kastner1996, Hora1999, Akras2017, Fang2018}, or from the outer regions of the PNe where the temperature is about 1000K \citep{Aleman2004, Matsuura2007b}.

\subsection{Dust and PAH Features}

The dust in SMP LMC 058 is carbon-rich. Amongst the most prominent features is the strong silicon carbide (SiC) emission at 11~$\mu$m and the rising continuum due to the thermal emission of warm dust. Strong emission features from PAHs also appear in the spectrum at 5.2, 5.7, 6.2, 7.7, 8.6, 11.2 and 12.7 ${\rm\mu m}$.

At sub-solar metallicities ($\sim0.2-0.5$ Z${_\odot}$), SiC is commonly observed in PNe, yet it is rarely seen in Galactic PNe or indeed during the earlier AGB evolutionary phase of metal-poor carbon stars \citep{Casassus2001, Zijlstra2006, Matsuura2007, Stanghellini2007, Bernard-Salas2008, Woods2011, Woods2012, Sloan2014, Ruffle2015, Jones2017b}.
The strength of the SiC flux in metal-poor PNe is highly sensitive to the radiation field \citep{BernardSalas2009}. This is likely due to a lower abundance of Si affecting the carbonaceous dust condensation sequence on the AGB. 
In this case, rather than SiC forming first, it instead forms in a mantle surrounding an amorphous carbon core \citep{Lagadec2007, Leisenring2008}. Then as the PNe dust becomes heated and photo-processed, the amorphous carbon evaporates increasing the SiC surface area and consequently, its feature strength, until a critical ionisation potential of $>$55 eV occurs at which point the SiC features disappear \citep{BernardSalas2009, Sloan2014}.

Dust is expected to undergo substantial evolution during the lifetime of a PN due to sputtering and grain-grain collisions which both `erode' and `shatter' the material \citep{Lenzuni1989, Jones1996}.  Coagulation through grain-grain collisions altering the grain size distribution may also be expected. However, the expected time scales for this dust evolution are unclear due to limited theoretical models of this stage. 
{\em Spitzer} IRS observations of SMP LMC 058 were obtained on 2005-05-21, 6284 days prior to our {\em JWST}/MRS observation of the source, thus enabling a temporal study of dust evolution in a PNe on a 17-year time scale. 
Following \citet{BernardSalas2009}, we measure the strength of the SiC feature by integrating the flux above a continuum-subtracted spectrum from 9 to 13.2 $\mu$m and then subtracting the flux contributions from the 11.2 $\mu$m PAH feature and the [Ne {\sc ii}] line. 
Due to the resolution of the MRS compared to the {\em Spitzer} spectra of SMP LMC 058, we detect several additional lines which contribute to the integrated flux in the SiC region; these lines include [S {\sc iv}] and H {\sc i}. Thus to obtain a reliable measurement of the SiC feature strength we also subtract the flux contribution from all emission lines in the 9 -- 13.2 $\mu$m region listed in Table~\ref{tab:lineIDandstrengths}.  A PAH feature at $\sim$12.6 $\mu$m likely contributes a small amount of flux to the measured SiC feature, however isolating and subtracting this emission contribution from the wing of the SiC feature is challenging even with the MRS spectral resolution. Additionally, an artefact at $\sim$12.2 $\mu$m due to a spectral leak \citep[e.g.,][]{Gasman2022} may also affect the integrated flux. 
Table~\ref{tab:PAHlinetrengths} gives the measured SiC centroid and corrected feature strength.  The latter agrees exceptionally well with the value of 29.72 $\pm$ 0.31 $\times10^{-16}$ $\mathrm{W\,m^{-2}}$ measured by \citet{BernardSalas2009} in the {\em Spitzer} data of SMP LMC 058. This suggests there is little to no evolution in the SiC dust on the 17-year time scales between the observations. Furthermore, the agreement between the measurements verifies the overall flux calibration of the MRS instrument \citep{Gasman2022}.


\begin{figure*}
\centering
\includegraphics[trim=0.05cm 0cm 0cm 0cm, clip=true,width=\hsize]{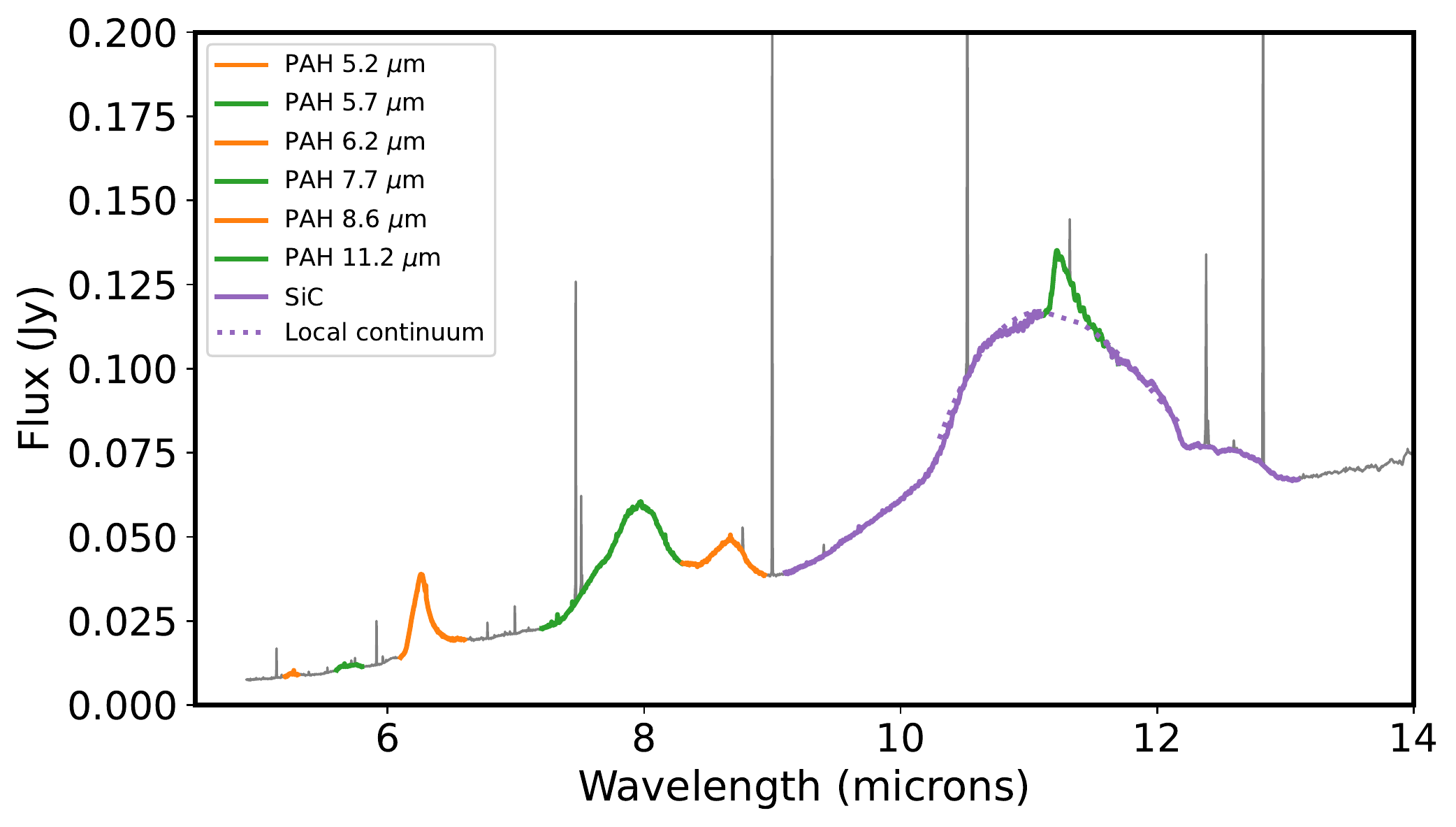}
\caption{The SiC and PAH features are highlighted in the spectra of SMP LMC 058. A local continuum fit to the 11.3 $\mu$m feature which is superimposed on the broad SiC emission feature is also shown. The colours highlight the spectral region for each feature, to which a local continuum was fit and the flux measured over.}
\label{Fig:PAH_measure}
\end{figure*}

In astronomical sources, the structure, wavelengths and relative strength of the PAHs can differ strongly between objects, with PNe showing the most pronounced variations in PAH profiles due to photoprocessing altering the ratio of aliphatics to aromatics \citep{Peeters2002, Pino2008, Matsuura2014, Sloan2014, Jensen2022}. 
Figure~\ref{Fig:PAH_measure} shows the PAHs in SMP LMC 058.
The PAHs in SMP LMC 058 are considered to have a class B profile by \citet{BernardSalas2009} and \citet{Sloan2014}. In this schema devised by \citet{Peeters2002} and \cite{vanDiedenhoven2004} the 6.2 PAH feature for class B objects has a peak between 6.24 and 6.28 $\mu$m; the dominant 7.7 PAH feature peaks between 7.8 to 8.0 $\mu$m; and the 8.6 PAH band is red-shifted. These values agree well with our measured centroids listed in Table~\ref{tab:PAHlinetrengths}.
Furthermore, the PAHs observed in SMP LMC 058 closely resemble those observed in the {\em ISO} SWS spectrum of the Galactic post-AGB star, HD 44179 (the Red Rectangle) which also shows strong aromatic features on top of a continuum \citep{Waters1998}.

The relative strength of the PAH features depends on a number of factors including the degree of ionisation of the radiation field \citep[e.g.,][]{Allamandola1999}. 
The strength of the PAH features in SMP LMC 058 was measured by integrating the flux of the feature above an adopted local continuum, fit to each side of the feature and measured using {\sc specutils} {\sc line\_flux}.  
Particular care was taken in fitting a continuum, too, and then measuring the 11.25 ${\rm\mu m}$ band (produced by the out-of-plane solo C–H bending mode) as this is superimposed on top of the broad SiC feature. Table~\ref{tab:PAHlinetrengths} presents the central wavelength of the features and the integrated flux.  
The ratio of the PAH strengths correlates with the source type and hence its physical conditions \citep{Hony2001}; ionized PAHs have strong features at 6.2, 7.7 and 8.6 $\mu$m whilst the 11.2 $\mu$m PAH feature is stronger for neutral PAHs. From the PAH line strengths given in Table~\ref{tab:PAHlinetrengths} it is evident that the 7.7$\mu$m feature dominates the total PAH emission.

\begin{table}
\caption{PAH and SiC Fluxes and Centroids.}
\label{tab:PAHlinetrengths}
\centering
\begin{tabular}{lccc}
\hline
\hline
 & Centroid & Integrated Flux & Integrated Flux Error  \\
 & $\mathrm{\mu m}$ & $\mathrm{W\,m^{-2}}$ & $\mathrm{W\,m^{-2}}$\\ 
\hline
PAH &    5.262   & 1.64$\times10^{-18}$ & 7.0$\times10^{-20}$ \\
PAH &    5.698   & 5.62$\times10^{-18}$ & 1.1$\times10^{-19}$ \\
PAH &    6.274   & 9.259$\times10^{-17}$ & 2.9$\times10^{-19}$ \\
PAH &    7.834   & 3.427$\times10^{-16}$ & 1.2$\times10^{-18}$ \\
PAH &    8.665   & 6.231$\times10^{-17}$ & 8.5$\times10^{-19}$ \\
PAH &    11.298  & 1.047$\times10^{-16}$ & 2.0$\times10^{-18}$ \\
SiC &    11.097  & 3.067$\times10^{-15}$ & 4.5$\times10^{-18}$ \\
\hline
\end{tabular}
\end{table}


Carbon-rich PNe can show a rich variety of solid-state material in their spectra in addition to PAHs.  The C$_{60}$ fullerene molecule typically exhibits features at $\sim$7.0, 8.5, 17.4 and 18.9 $\mu$m, and all four were first identified in the spectrum of the Galactic PN TC-1 \citep{Cami2010}.  Fullerenes have since been detected in several other PNe \citep[e.g.,][]{GarciaHernandez2010, GarciaHernandez2011, Sloan2014}.  The still-unidentified 21~$\mu$m emission feature, first detected by \citealt{Kwok1989}, can also appear in carbon-rich PNe, often associated with unusual PAH emission and aliphatic hydrocarbons \citep{Cerrigone2011, Matsuura2014, Sloan2014, Volk2020}.
The spectra of SMP LMC 058 from the IRS on Spitzer did not show any of these unusual hydrocarbon-related features, but the improved spectral resolution of the MRS allows for a much more careful examination.  Nonetheless, these additional features remain too weak to be detected.  SMP LMC 058 presents a classic Class B PAH spectrum, as expected for objects which have evolved to the young PN stage \citep{Sloan2014}.  Younger objects which could still be described as post-AGB objects would show the 21~$\mu$m feature and/or aliphatics. \cite{Sloan2014} identified SMP LMC 058 as a member of the Big-11 group because of the combination of a strong SiC emission feature and the 11.2~$\mu$m PAH feature and the absence of fullerenes. They also noted that the Big-11 group is related to the PNe that show fullerenes, and that the presence or absence of fullerenes may be due to something as simple as which have a clear line of sight to the interior of the dust shells where the fullerenes are expected to be present.

\section{Summary and Conclusions}
\label{Summ_Conc}

We have presented MIRI/MRS spectra of the carbon-rich planetary nebula SMP LMC 058 located in the Large Magellanic Cloud. SMP LMC 058 is a point source in the MRS data and its spectrum contains the only spatially and spectrally unresolved emission lines observed during the commissioning of the {\em JWST} Medium-Resolution Spectrometer.
In the MRS spectrum, we detected 51 emission lines, of which 47 were previously undetected in this source. 
The strongest emission lines were used to determine the spectral resolutions of the MIRI MRS instrument.  The resolving power is R $>$  3660 in channel 1, R $>$ 3430 in channel 2, R $>$ 3200 in channel 3, and R  $>$ 1920 in channel 4. 
This on-sky performance is comparable to the resolution determined from the ground calibration of the MRS which provides resolving powers from 4000 at channel 1 to 1500 at channel 4. Furthermore, a comparison of the line strengths and spectral continuum to previous observations of SMP LMC 058 with the IRS on the {\em Spitzer} was used to verify the absolute flux calibration of the MRS instrument. 
The MRS spectra confirm that the carbon-rich dust emission is from grains and not isolated molecules and that there is little to no time evolution of the SiC dust and emission line strengths in the 17 years between the observations. 
The PAH emission is dominated by the 7.7$\mu$m feature. The strong PAHs and SiC in the spectra are consistent with the lack of high-excitation lines detected in the spectra, which if present, would indicate a hard radiation field that would likely destroy these grains. 
These commissioning data reveal the great potential and resolving power of the MIRI MRS to study line, molecular and solid-state features in individual sources in nearby galaxies.

\section*{Acknowledgements}

We thank the referee Jeronimo Bernard-Salas for the constructive report and useful suggestions that improved our manuscript. 
We thank Kay Justtanont and Kathleen Kraemer for their insights, comments and discussions.
This work is based on observations made with the NASA/ESA/CSA James Webb Space Telescope. The data were obtained from the Mikulski Archive for Space Telescopes at the Space Telescope Science Institute, which is operated by the Association of Universities for Research in Astronomy, Inc., under NASA contract NAS 5-03127 for JWST. These observations are associated with program \#1049. This work is based in part on observations made with the Spitzer Space Telescope, which was operated by the Jet Propulsion Laboratory, California Institute of Technology under a contract with NASA

O.C.J acknowledge support from an STFC Webb fellowship.
J.A.-M. and A.L acknowledge support by grant PIB2021-127718NB-100 from the Spanish Ministry of Science and Innovation/State Agency of Research MCIN/AEI/10.13039/501100011033 and by “ERDF A way of making Europe”.
P.J.K acknowledges financial support from the Science Foundation Ireland/Irish Research Council Pathway programme under Grant Number 21/PATH-S/9360.
I.A., D.G., and B.V. thank the European Space Agency (ESA) and the Belgian Federal Science Policy Office (BELSPO) for their support in the framework of the PRODEX Programme. PG would like to thank the University Pierre and Marie Curie, the Institut Universitaire de France, the Centre National d'Etudes Spatiales (CNES), the "Programme National de Cosmologie and Galaxies" (PNCG) and the "Physique Chimie du Milieu Interstellaire" (PCMI) programs of CNRS/INSU, with INC/INP co-funded by CEA and CNES,  for there financial supports.

MIRI draws on the scientific and technical expertise of the following organisations: Ames Research Center, USA; Airbus Defence and Space, UK; CEA-Irfu, Saclay, France; Centre Spatial de Liége, Belgium; Consejo Superior de Investigaciones Científicas, Spain; Carl Zeiss Optronics, Germany; Chalmers University of Technology, Sweden; Danish Space Research Institute, Denmark; Dublin Institute for Advanced Studies, Ireland; European Space Agency, Netherlands; ETCA, Belgium; ETH Zurich, Switzerland; Goddard Space Flight Center, USA; Institute d'Astrophysique Spatiale, France; Instituto Nacional de Técnica Aeroespacial, Spain; Institute for Astronomy, Edinburgh, UK; Jet Propulsion Laboratory, USA; Laboratoire d'Astrophysique de Marseille (LAM), France; Leiden University, Netherlands; Lockheed Advanced Technology Center (USA); NOVA Opt-IR group at Dwingeloo, Netherlands; Northrop Grumman, USA; Max-Planck Institut für Astronomie (MPIA), Heidelberg, Germany; Laboratoire d’Etudes Spatiales et d'Instrumentation en Astrophysique (LESIA), France; Paul Scherrer Institut, Switzerland; Raytheon Vision Systems, USA; RUAG Aerospace, Switzerland; Rutherford Appleton Laboratory (RAL Space), UK; Space Telescope Science Institute, USA; Toegepast- Natuurwetenschappelijk Onderzoek (TNO-TPD), Netherlands; UK Astronomy Technology Centre, UK; University College London, UK; University of Amsterdam, Netherlands; University of Arizona, USA; University of Bern, Switzerland; University of Cardiff, UK; University of Cologne, Germany; University of Ghent; University of Groningen, Netherlands; University of Leicester, UK; University of Leuven, Belgium; University of Stockholm, Sweden; Utah State University, USA. A portion of this work was carried out at the Jet Propulsion Laboratory, California Institute of Technology, under a contract with the National Aeronautics and Space Administration.

The following National and International Funding Agencies funded and supported the MIRI development: NASA; ESA; Belgian Science Policy Office (BELSPO); Centre Nationale d'Etudes Spatiales (CNES); Danish National Space Centre; Deutsches Zentrum fur Luftund Raumfahrt (DLR); Enterprise Ireland; Ministerio De Economia y Competividad; Netherlands Research School for Astronomy (NOVA); Netherlands Organisation for Scientific Research (NWO); Science and Technology Facilities Council; Swiss Space Office; Swedish National Space Agency; and UK Space Agency.

\

\noindent {\it Facilities:} {\em JWST} (MIRI/MRS) - James Webb Space Telescope.


\section*{Data availability}

JWST data were obtained from the Mikulski Archive for Space Telescopes at the Space Telescope Science Institute (\href{https://archive.stsci.edu/}{https://archive.stsci.edu/}). 



\bibliographystyle{mnras}

\bibliography{main} 







\bsp	
\label{lastpage}
\end{document}